\documentclass[a4paper,11pt]{article}
\usepackage{epsfig}
\usepackage{color}
\usepackage{amsmath,empheq}
\usepackage{mathtools}
\usepackage{amssymb}
\usepackage{hyperref}
\usepackage{enumerate}
\usepackage{footmisc}
\usepackage{lineno}

\usepackage{graphicx}

\begin{document}

\title{Alteration of cerebrovascular haemodynamic patterns due to atrial fibrillation: an \emph{in silico} investigation}
\date{}
\maketitle
\author{S. Scarsoglio$^{1}$\footnote[1]{Corresponding author: stefania.scarsoglio@polito.it}, A. Saglietto$^{2}$, M. Anselmino$^{2}$, F. Gaita$^{2}$ and L. Ridolfi$^{3} \vspace{+0.5cm}$ \\
\it $^{1}$ Department of Mechanical and Aerospace Engineering, Politecnico di Torino, Torino, Italy \smallskip\\
\it $^{2}$ Division of Cardiology, Department of Medical Sciences, "Citt\`{a} della Salute e della Scienza" Hospital, University
of Turin, Torino, Italy \smallskip\\
\it $^{3}$ Department of Environmental, Land and Infrastructure Engineering, Politecnico di Torino, Torino, Italy}


\begin{abstract}
There has recently been growing evidence that atrial fibrillation (AF), the most
common cardiac arrhythmia, is independently associated with the risk of
dementia. This represents a very recent frontier with high social impact for
the number of individuals involved and for the expected increase in AF incidence
in the next 40 years. Although a number of potential haemodynamic
processes, such as microembolisms, altered cerebral blood flow, hypoperfusion
and microbleeds, arise as connecting links between the two pathologies, the
causal mechanisms are far from clear. An \emph{in silico} approach is proposed that
combines in sequence two lumped-parameter schemes, for the cardiovascular
system and the cerebral circulation. The systemic arterial pressure is obtained
from the cardiovascular system and used as the input for the cerebral circulation,
with the aim of studying the role of AF on the cerebral haemodynamics
with respect to normal sinus rhythm (NSR), over a 5000 beat recording. In particular,
the alteration of the haemodynamic (pressure and flowrate) patterns in
the microcirculation during AF is analysed by means of different statistical
tools, from correlation coefficients to autocorrelation functions, crossing
times, extreme values analysis and multivariate linear regression models.
A remarkable signal alteration, such as a reduction in signal correlation
(NSR, about 3 s; AF, less than 1 s) and increased probability (up to three to
four times higher in AF than in NSR) of extreme value events, emerges for
the peripheral brain circulation. The described scenario offers a number of
plausible cause-effect mechanisms that might explain the occurrence of critical
events and the haemodynamic links relating to AF and dementia.

\end{abstract}

\section{Introduction}

Atrial fibrillation (AF), leading to an irregular and faster heart rate, is the most
common tachyarrhythmia with an estimated number of 33.5 million individuals
affected worldwide in 2010 \cite{Piccini_2014}, and its incidence is expected to double within the
next 40 years \cite{Lloyd-Jones_2010}. Besides thromboembolic transient ischaemic attack (TIA) and
stroke—whose risk is increased fivefold in patients with AF \cite{Wolf_1991} and is associated
with both cerebral impairment and dementia \cite{Zhu_1998} - it has been recently observed
that AF is independently associated with cognitive decline through a range of
different potential haemodynamic mechanisms, such as silent cerebral infarctions
(SCIs) as a result of microembolization \cite{Kalantarian_2014,Gaita_2013}, altered cerebral blood flow \cite{Sabatini_2000}, hypoperfusion \cite{Jacobs_2014} and microbleeds, whose repetition increases the risk of intracerebral haemorrhagic events and dementia by five times \cite{Kanmanthareddy}.

Although representing a currently debated topic \cite{Dixit_2015}, there is growing evidence that AF - independently of clinically relevant events - enhances the risk of dementia and cognitive deficit \cite{Jacobs_2014,Kanmanthareddy,Alonso_2016}. Several different kinds of observational works - such as meta-analyses \cite{Kalantarian_2013}-\cite{Kwok_2011}, reviews \cite{Hui_2015,Udompanich_2013}, cross-sectional \cite{Ott_1997}-\cite{Chen_2016}, cohort and longitudinal \cite{Chen_2014}-\cite{Marzona_2012} studies - confirm an independent association between AF and cognitive decline at differing grades of severity. Only a few with critical limiting aspects and potential sources of bias, such as small population \cite{O'Connell_1998}, very high rate of loss during follow-up \cite{Park_2007} and very elderly subjects (aged 85 and older) \cite{Rastas_2007}, found no significant relation between AF and cognitive impairment.

However, most of the above observational studies can only show an association between AF and cognitive impairment and not a causal relation based on haemodynamics for any of the known potential mechanisms. Recently, the role of SCIs in cognitive function during AF has been assessed through magnetic resonance imaging \cite{Gaita_2013}. Although some of the haemodynamic consequences of AF, such as lower diastolic cerebral perfusion and decreased blood flow in the intracranial arteries, have been reported ( \cite{Jacobs_2014,Kanmanthareddy,Hui_2015}and references therein), the linking mechanisms with cognitive impairment remain theoretical or mainly undetermined.

\noindent To the best of our knowledge, the specific impact of the altered AF heart rate on the cerebral haemodynamics is still in great part unexplored. In fact, currently adopted clinical techniques in the field of cerebral haemodynamics - such as transcranial doppler ultrasonography - lack the resolving power to give any insights on the micro-vasculature, in terms of flow and pressure signals. In particular, little is known about the consequences of AF treatment on the evolution of cognitive decline. So far, only a few studies have examined the potential benefits from AF treatment in reducing cognitive impairment. Increased cognitive dysfunction was found to relate to less effective oral anticoagulation treatment \cite{Flaker_2010}, while AF patients who underwent catheter ablation had a lower risk of dementia than those who did not \cite{Bunch_2011}. These studies, though not prospective and with biased information, give insights that specific treatments for AF could modify the risk of dementia. The intriguing recent hints offered by literature encourage a deeper comprehension of the AF effects on hypoperfusion and irregular cerebral blood flow, which is still lacking \cite{Kalantarian_2016}.

The efficiency of a mathematical modelling approach to describe the cerebral circulation has been widely recognized and \emph{in silico} haemodynamics is currently a rising field of research, e.g. \cite{Sadraie_2014,Campodeano_2015}. In a previous work, we obtained the first exploratory results by adopting two lumped-parameter models for the cardiovascular and cerebral circuits, which highlighted the onset of critical events - such as hypoperfusions and hypertensive events - at the arteriolar and capillary levels during AF \cite{SciRep_2016}. Aim of the present work is to understand and analyze - through a systematic and extensive signal analysis - possible haemodynamic-based causal relations underlying the occurrence of such critical events, for which AF may imply cognitive dysfunction. The statistical tools exploited here are borrowed from classical time-series analysis and include cross-correlation functions between the input pressure/flow rate signals and the corresponding downstream signals, auto-correlation functions in different cerebral regions, distribution of consecutive time lapses spent above/below the mean value of the pressure and flow rate temporal signals, detection of minimum and maximum haemodynamic values per beat, quadrant analysis, and multivariate linear regression models for the haemodynamic variables (averaged by beat). A model-based estimation of these critical events can offer useful hints for the assessment of some of the AF treatments, in particular rate and rhythm control strategies, as it can suggest priority treatment to minimize neurodegenerative changes.

\noindent To isolate single cause-effect relations and ascertain which AF-driven variation mostly affects the cerebral circulation and should  therefore be taken under strict control, a comparative signal analysis (in terms of pressure and flow rate time-series) is proposed between normal sinus rhythm (NSR) and AF signals over a 5000 beat recording. The paper is organized as follows. In the Materials and Methods Section the stochastic modelling, consisting of the cardiovascular and cerebral systems, is introduced. The following section (Pressure and Flow Rate Signal Analysis) proposes a collection of different statistical tools to carry out a systematic study of the signal variation. In the Discussion, a summarizing framework explaining the reasons for AF-induced changes in the micro-circulation is given. Conclusions remark that the emergence of critical events in AF turns out to be caused by the signal alteration - especially in terms of correlation, memory and complexity - induced by AF in the micro-vascular haemodynamics.

\section{Materials and Methods}

\subsection{Computational modelling and beating features}

The stochastic modelling of the AF-induced cerebral haemodynamics has been recently proposed \cite{SciRep_2016} and consists of three sequential steps. Figure 1 describes the modelling process adopted (panels 1, 2, 3) and shows a representative pressure time series obtained as outputs (panel 4). The proposed stochastic algorithm combines two different lumped models in sequence: the cardiovascular model is exploited to obtain the systemic arterial pressure, $P_a$, which is then used as the forcing input for the next cerebral model.

\begin{figure}
\hspace{-3.3cm}
\includegraphics[width=1.5\columnwidth]{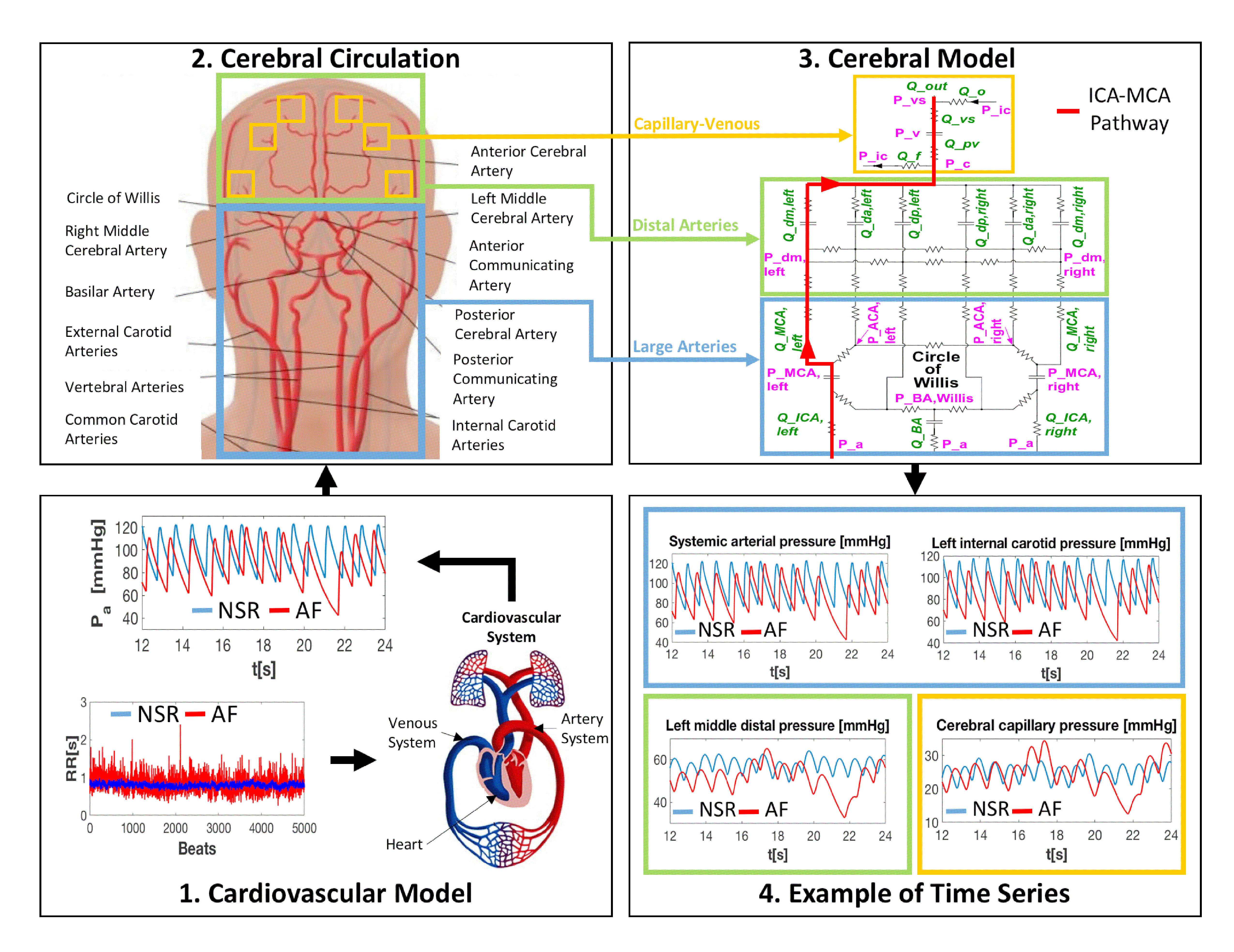}
\caption{Scheme of the \emph{in silico} approach (the figure should be read clockwise starting from the left bottom panel 1). (1) Cardiovascular model: 5000 extracted RR records in NSR (blue) and AF (red), and examples of $P_a$ time series obtained through the cardiovascular model. (2) Cerebral circulation: sketch of the cerebral vasculature forced by the $P_a$ input, which is obtained from the cardiovascular model described in panel 1. (3) Cerebral mathematical model:  $R$, resistance; $C$, compliance; $Q$, flow rate; $P$, pressure. The left ICA-MCA pathway is highlighted in red and is composed of $P_a$, $Q_{ICA,left}$, $P_{MCA,left}$, $Q_{MCA,left}$, $P_{dm,left}$, $Q_{dm,left}$, $P_c$ and $Q_{pv}$. (4) Examples of pressure time series. Representative resulting time series for the pressure along the ICA-MCA pathway, in NSR (blue) and AF (red) conditions, obtained from the cerebral model described in panel 3. In panels 2, 3, 4, the coloured boxes refer to different cerebral regions: large arteries (light blue), distal arteries (green), capillary/venous circulation (yellow).}
\end{figure}

\begin{itemize}
\item \emph{Building the RR intervals.} We recall that $RR$ [s] is the temporal interval between two consecutive heart beats, while the heart rate, $HR$, is the number of heart beats per minute. Normal sinus and fibrillated beating are modelled via artificially built $RR$ intervals based on NSR and AF beating features (see the details in \cite{MBEC_2014}). Normal $RR$ heart beats are extracted from a correlated pink Gaussian distribution (mean $\mu=0.8$ s, standard deviation, $\sigma=0.06$ s), which is the typical distribution observed during sinus rhythm for $RR$ \cite{MBEC_2014}. AF beatings are instead extracted from an exponentially Gaussian modified (EGM) distribution (mean $\mu=0.8$ s, standard deviation $\sigma=0.19$ s, rate parameter $\gamma=7.24$ Hz), which is unimodal and represents the most common AF distribution (60-65\% of the cases) \cite{Hennig_2006,Hayano_1997}. The exponential contribution is responsible for the uncorrelated nature of the AF beating. Comparison between NSR and AF is proposed at the same mean heart rate (75 beats per minute (b.p.m.)) to facilitate analysis of the results. 5000 beats are extracted and then simulated for both configurations in order to achieve the statistical stationarity for the main statistics of the outcomes (the 5000 RR beats extracted in NSR and AF conditions are reported in the first panel of Fig. 1).
\item \emph{Cardiovascular model.} Once the $RR$ extraction is complete, a lumped cardiovascular model is used to obtain the systemic arterial pressure ($P_a$). The model was first proposed \cite{Shi_2006} to describe, through a Windkessel approach, the complete cardiovascular system. It includes the  systemic and  venous  circuits  together  with  the four cardiac chambers which are actively modelled. By means of a network of compliances,  resistances, and inductances, the cardiovascular dynamics is expressed in terms of pressures, flow rates, volumes, and valve opening angles. After being validated during resting conditions over more than 30 clinical datasets \cite{MBEC_2014,CMBBE_2016}, the model has been exploited to study left valvular diseases \cite{PeerJ_2016} during AF and the effect of increased heart rate in resting conditions \cite{PlosOne_2015} and under exercise \cite{PlosOne_2017}. To account for AF conditions, both atria are considered as passive (while in NSR they actively contract). We point out that the cardiovascular and cerebral models are combined in sequence: once the systemic arterial pressures, $P_a$, are obtained from the cardiovascular scheme in NSR and AF conditions, they are then used as forcing inputs for the cerebral model. Examples of $P_a$ time series are reported in the first panel of Fig. 1.
\item \emph{Cerebral model.} Zero-dimensional modelling for the cerebrovascular dynamics has been proposed \cite{Ursino} to study the whole (arterial and venous) cerebral circulation (Panels 2 and 3, Fig. 1). Similarly to the cardiovascular model, a framework of resistances ($R$, [mmHg s/ml]) and compliances ($C$, [ml/mmHg]) accounts for the dissipation effects and the elastic properties of vessels, respectively. The cerebral circulation is expressed in terms of pressure ($P$, [mmHg]), volume ($V$, [ml]), flow rate ($Q$, [ml/s]), and can be divided into three principal regions: large arteries, distal arterial circulation, capillary-venous circulation. The first section is composed of the afferent arteries and the circle of Willis, while the six main cerebral arteries link this region to the downstream distal circuit. The distal arterial circulation includes the pial circulation and the intracerebral arteries-arterioles, and is split into six regional districts, independently controlled by autoregulation and $CO_2$ reactivity. The cerebrovascular control mechanisms are individually described by means of first-order low-pass dynamics, acting to directly maintain the physiological flow rate level. The consequent autoregulation mechanisms of vasodilatation and vasoconstriction are ruled by a temporal variation of the distal compliances, $C$, and resistances, $R$. A unique pressure downstream from the distal region represents the capillary pressure. The cerebral venous circulation is defined by two-element Windkessel modelling, while the cerebrospinal fluid circulation is formed at the level of cerebral capillaries. In the following, a single pathway (internal carotid artery (ICA) -- middle cerebral artery (MCA)) highlighted in Fig. 1 (Panel 3) is studied as representative of the blood flow and pressure distributions from large arteries to the capillary-venous circulation: left internal carotid artery ($P_a$ and $Q_{ICA,left}$), middle cerebral artery ($P_{MCA,left}$ and $Q_{MCA,left}$), middle distal region ($P_{dm,left}$ and $Q_{dm,left}$), capillary-venous circulation ($P_c$ and $Q_{pv}$). Examples of the pressure time series of the ICA-MCA pathway are shown in Fig. 1, Panel 4. More details on the cerebral model are offered elsewhere \cite{SciRep_2016}.
\end{itemize}

\section{Pressure and Flow Rate Signal Analysis}

The analysis, involving a record of 5000 beats (for both NSR and AF), focuses on the pressure and flow rate time series along the ICA-MCA pathway and can be divided in two main parts: (i) analyses of the continuous time series and (ii) beat-by-beat analyses. In the first set, the signal is continuous and defined by all the temporal instants of the time series. In the second set, the signal is discretized and one per beat data are obtained. Therefore, discrtized time series are composed of 5000 elements, corresponding to the 5000 beats simulated. The \emph{i-th} element may contain the average values ($\overline{Q}$ and $\overline{P}$), as well as the maximum ($Q_{max}$ and $P_{max}$) and minimum ($Q_{min}$ and $P_{min}$) values of the related haemodynamic variables computed over the \emph{i-th} beat.

\subsection{Complete time series analysis}

\subsubsection{Linear correlation coefficient and auto-correlation function}

\begin{table}[h!]
\centering
\begin{tabular}{|c|c|c|}
  \hline
   & NSR & AF \\
   \hline
  $(P_a, P_{MCA,left})$ & 1.00  & 1.00  \\
   \hline
  $(P_a, P_{dm,left})$ & 0.83  & 0.76  \\
   \hline
  $(P_a, P_{c})$ & 0.83  & 0.65  \\
   \hline
  $(Q_{ICA,left}, Q_{MCA,left})$ &  0.99  & 0.98  \\
   \hline
  $(Q_{ICA,left}, Q_{dm,left})$ &  0.87  & 0.72  \\
   \hline
  $(Q_{ICA,left}, Q_{pv})$ &  0.88  & 0.80  \\
  \hline
\end{tabular}
\label{correlation_table}
\caption{Linear correlation coefficient, $\rho$, for NSR and AF conditions. Signals are normalized with respect to their mean and standard deviation values, as follows: $x_n= (x - \mu_x)/\sigma_x$. For mean and standard deviation values, please refer to \cite{SciRep_2016}.}
\end{table}

The linear correlation coefficient is calculated between the signal entering into the brain ($P_a$) and the signals downstream up to the capillary region ($P_{MCA,left}$, $P_{dm,left}$, $P_c$). Analogous computation is performed for flow rates, that is, $Q_{ICA,left}$ with respect to $Q_{MCA,left}$, $Q_{dm,left}$ and $Q_{pv}$. In Table 1 the linear correlation coefficient, $\rho$, between couples of haemodynamic signals is reported in NSR (II column) and AF (III column) conditions. In both NSR and AF conditions the correlation - which remains very high in the middle cerebral artery section - is damped towards the distal circulation. However, the damping is by far more relevant in the fibrillated condition. At the capillary-venous level, the correlation in AF is decreased by up to $21\%$ with respect to NSR for the pressure, while in the distal region is decreased by up to $17\%$ for the flow rate. The key aspect emerging here is that AF haemodynamic signals in the deep cerebral circulation are more prone than the corresponding NSR signals to lose their temporal interdependence with respect to the large artery circulation. Peripheral signals differ much more from the corresponding input signals during AF than during NSR.

\begin{figure}
\includegraphics[width=\columnwidth]{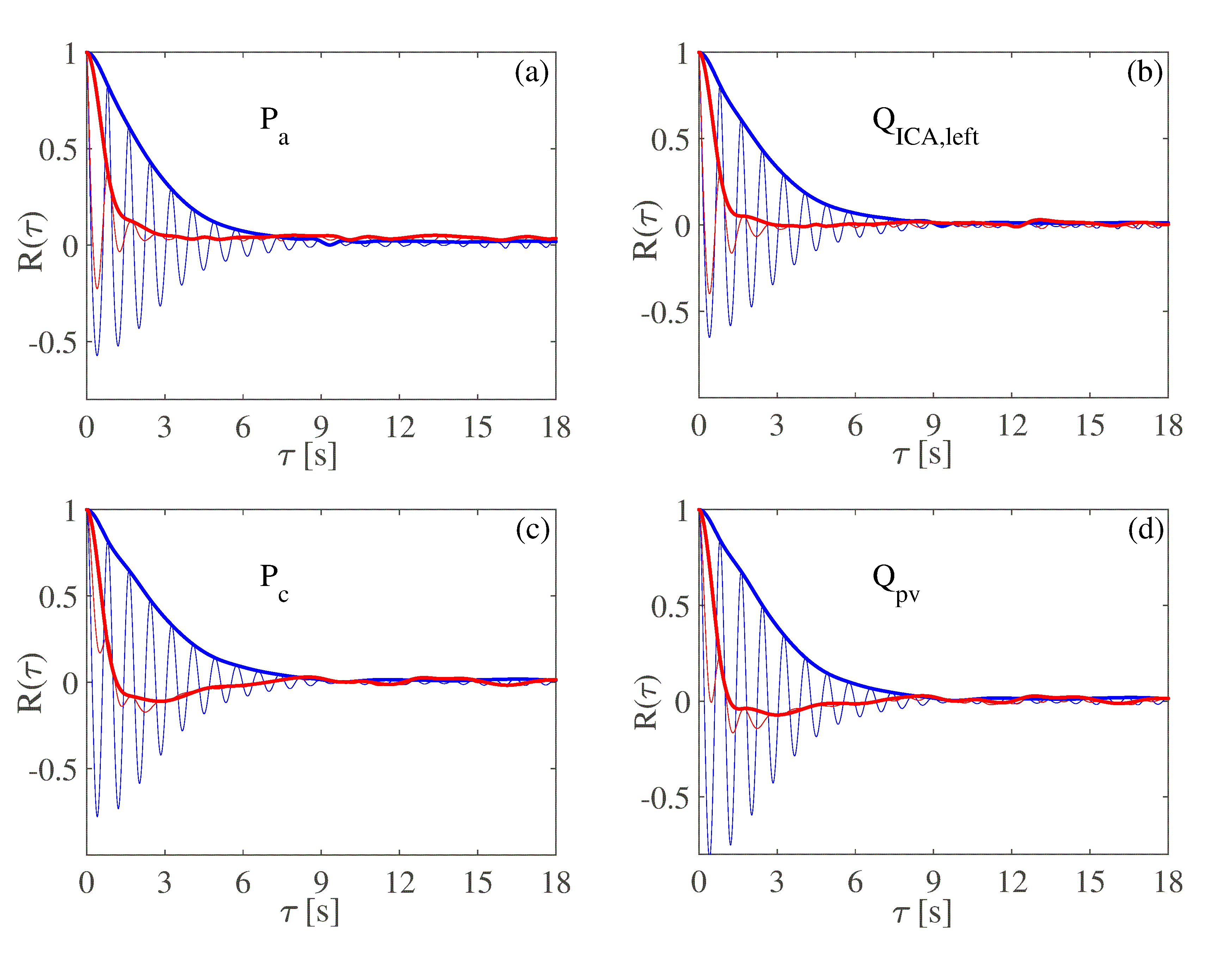}
\caption{Autocorrelation functions (thin curves), $R(\tau)$, and corresponding envelopes (thick curves), $R_{env}(\tau)$, as functions of the delay time, $\tau$. Pressure (left, panels a and c) and flow rate (right, panels b and d) signals from large arteries (top, panels a and b) to the capillary/venous region (bottom, panels c and d). NSR: blue, AF: red.}
\end{figure}

\noindent Fig. 2 shows the autocorrelation functions, $R(\tau)$, together with the corresponding envelopes, $R_{env}(\tau)$, of pressures and flow rates at the large arteries level (Fig. 2a,b) and in the capillary-venous region (Fig. 2c,d) for both NSR (blue) and AF (red) (for details, see the Appendix A). NSR autocorrelations in all cerebral regions display quasi-repetitive patterns (with period of about 0.8 s), and a decay in amplitude over the delay axis. The coherence times reported in Appendix A provide evidence that NSR signals maintain long-term memory (around about 4 beats) through the whole ICA-MCA pathway (temporal coherence even increases a bit towards the distal/capillary circulation). The picture is substantially different in AF. For the input signals, $R(\tau)$ still shows a remaining quasi-periodicity although the decay rate is very high. In the capillary region, the decrease in $R(\tau)$ resembles the behaviour of random signals. The coherence times, $\tau_c$, in the AF regime confirm a loss of memory, with values varying between 1.11 and 0.87 s. Towards the deep cerebral circulation the signal memory deteriorates even more so that capillary/venous haemodynamic signals during AF reveal short-time memory features ($\tau_c<1$ s). Therefore, the haemodynamic signals exhibit much more complexity and random-like features approaching the deep circulation in AF than in NSR.

\subsubsection{Crossing time analysis}

\begin{figure}
\includegraphics[width=\columnwidth]{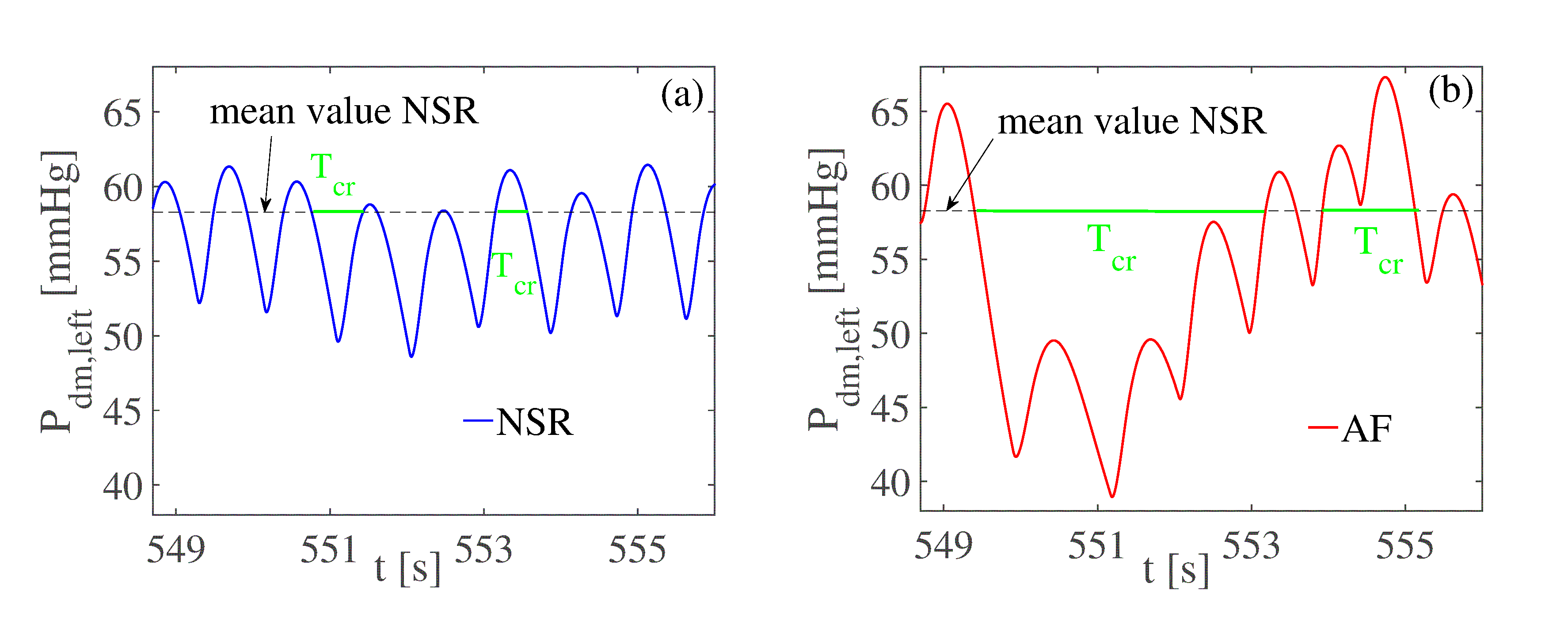}
\caption{Examples of $T_{cr}$ evaluation for an illustrative portion of the $P_{dml,left}$ time series (a: NSR, b: AF). $T_{cr}$ intervals are indicated in green.}
\end{figure}

Quantification of the consecutive time lapses spent by each variable above or below a certain threshold is introduced here through the crossing time, $T_{cr}$: it represents the temporal interval spent by the haemodynamic variable above or below the threshold indicated by the mean value in NSR. Fig. 3 (panels a and b) displays representative examples of $T_{cr}$ intervals for $P_{dm,left}$, during NSR and AF. The $T_{cr}$ intervals are found throughout the whole temporal series to evaluate how AF influences the duration of excursions from the reference mean value in NSR.

\begin{figure}
\includegraphics[width=\columnwidth]{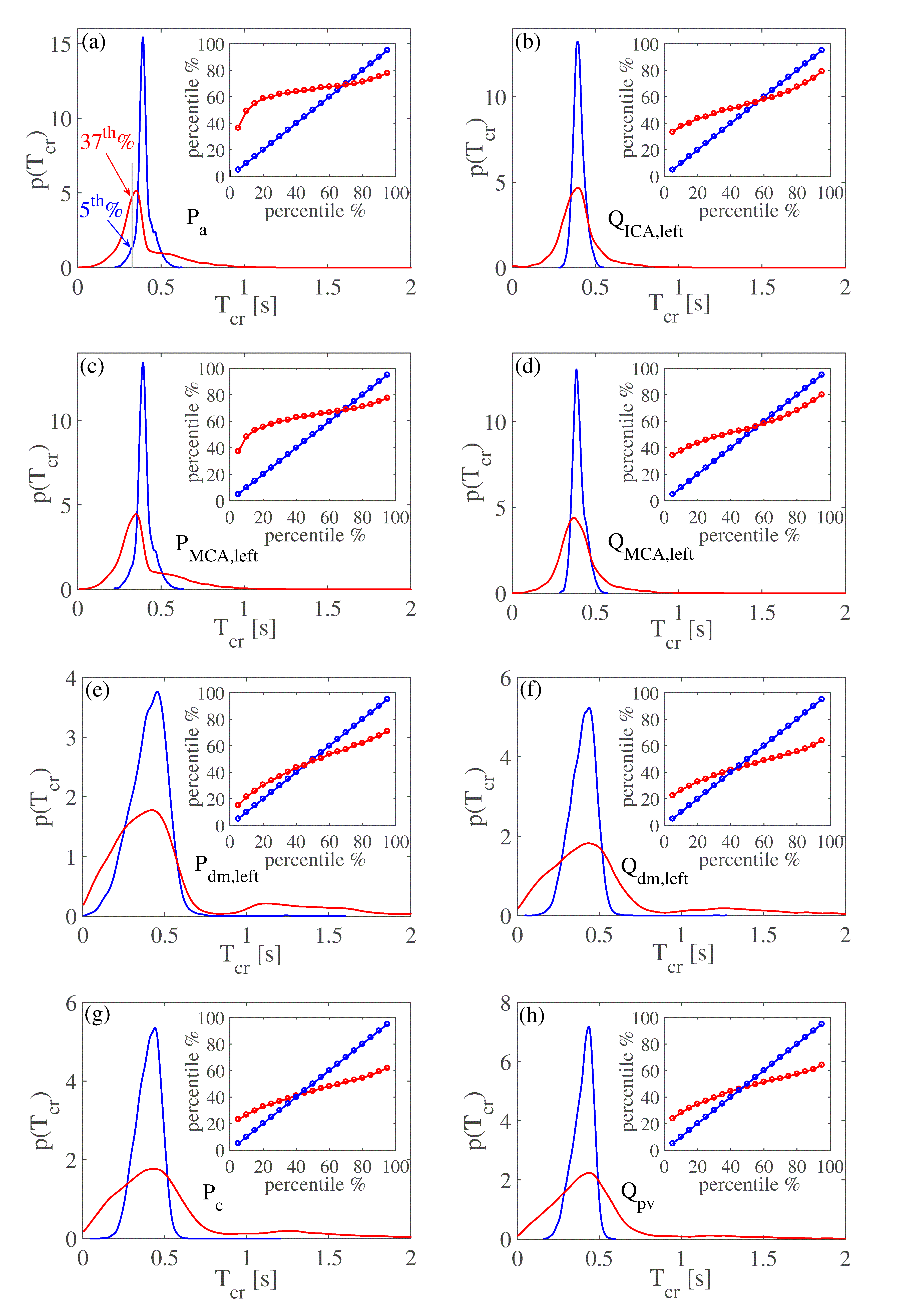}
\caption{PDFs of the crossing times, $T_{cr}$: NSR (blue), AF (red). Inserts show the percentile analysis of the $T_{cr}$ values, demonstrating to which AF percentile each NSR percentile corresponds.}
\end{figure}

As the crossing times, $T_{cr}$, are computed over the whole temporal series, we can then evaluate their probability density functions (PDFs) in NSR and AF (Fig. 4). During NSR in the large artery region (blue curves, panels from a to d), $T_{cr}$ values are narrowly centered around the mean value, which is half of the average beat, i.e. 0.4 s, thereby showing a stable oscillation of the signals around their mean values. Going towards the distal/capillary region (blue curves, panels from e to h), the mean values do not substantially vary, but the variability around them increases, revealing wider probability density functions. In AF conditions, in the large artery region (red curves, panels from a to d), the mean values are comparable to those observed during AF, while the standard deviation values are increased by 3 to 4 times. In the deep circulation (red curves, panels from e to h), the $T_{cr}$ mean values increase with respect to NSR and the standard deviation values increase by up to 3-4 times with respect to the AF large artery region. The PDFs display much more pronounced right tails and lose the symmetry shown during NSR. As displayed in the example of Fig. 3, the increased importance of the right-tails implies that the AF signals lose their periodicity around the mean value and spend long times (up to 2-3 s) consecutively well above or below the physiological threshold, without crossing it.

\noindent Complementary information is related to the percentile analysis. To highlight the AF-induced changes at the cerebral level, during NSR we computed the percentiles, from the 5th to the 95th (separated by 5ths), of the different quantities analyzed, conferring on these percentiles the role of reference NSR thresholds. In AF, we then evaluated to which percentile each of the nineteen NSR thresholds corresponds, thus quantifying how AF modifies the probability of reaching extreme values. An example is reported in Fig. 4, panel a: the $T_{cr}$ value indicated by the 5th percentile in NSR corresponds to the 37th percentile in AF. This means that a value which is extremely low and rarely reached in NSR becomes common and frequently attained in AF.

\noindent The inserts in Fig. 4 report the percentile analysis performed on the $T_{cr}$ values and represent to which AF percentile (red) each NSR percentile (blue) corresponds. It can be noted that in the large artery region low NSR percentiles (5-20 $\%$) correspond to quite high AF percentiles (35-50 $\%$), especially for the pressure. This means that shorter $T_{cr}$ values are more likely to occur in AF than in NSR. This picture
does not apply to the distal/capillary circulation, where, in addition, high NSR percentiles (95-80 $\%$) relate to much lower AF percentiles (70-50 $\%$). Thus, in the deep circulation, $T_{cr}$ values are statistically longer in AF than in NSR. This scenario is similarly observed for both pressure and flow rate and demonstrates in AF an increased probability of having long temporal ranges where excursions from the baseline haemodynamic value can develop.

\noindent The combined analysis of PDFs and percentile variation of the crossing time $T_{cr}$ reveals in the distal cerebral region a higher probability of extreme value events, such as hypoperfusions or hypertensive peaks, since the pressure and flow rate signals remain above or below their reference values for much longer (and consecutively). With the following beat-by-beat analysis we will be able to specify which kind of critical events may emerge, whether below (hypo) or above (hyper) the NSR haemodynamic thresholds.

\subsection{Beat-by-beat analysis}

\subsubsection{Minimum and maximum values analysis}

Minimum ($Q_{min}$ and $P_{min}$) and maximum ($Q_{max}$ and $P_{max}$) values over a cardiac beat are considered here, recalling that these are instantaneous haemodynamic values. In Appendix B, the mean and standard deviation values of the 5000 minimum and maximum values are reported for the haemodynamic variables. Fig. 5 presents the PDFs of the minimum and maximum values for pressures and flow rates at the large arteries level (Fig. 5a,b) and in the capillary-venous region (Fig. 5c,d) in NSR (blue) and AF (red). In NSR maximum and minimum PDFs are narrowly centered around the relative mean values and the coefficients of variation ($c_v=\sigma/\mu$) are well below 0.1, with values which do not significantly vary along the ICA-MCA pathway. In AF, the mean values do not essentially vary with respect to NSR (apart from $P_a$), while standard deviation values are significantly larger, leading to $c_v$ values often above 0.1.

\begin{figure}[h!]
\includegraphics[width=\columnwidth]{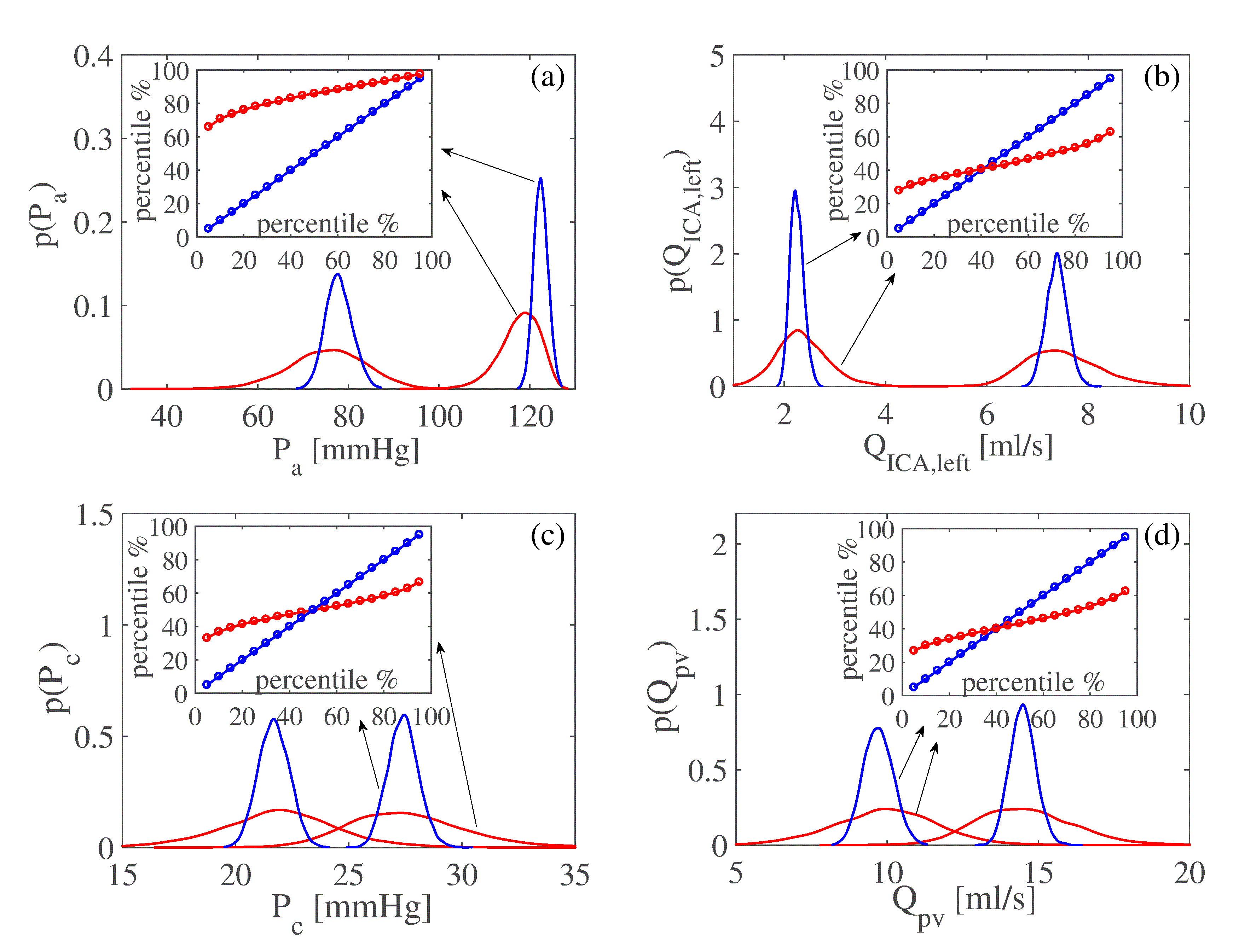}
\caption{Probability density functions of the maximum ($P_{max}$ and $Q_{max}$) and minimum ($P_{min}$ and $Q_{min}$) values, NSR: blue, AF: red. Pressure (left, panels a and c) and flow rate (right, panels b and d) from large arteries (top, panels a and b) to the capillary/venous region (bottom, panels c and d). Inserts represent the percentile analysis, showing how percentiles in NSR are modified in AF. Percentiles of maxima are reported for pressures, while percentiles of minima are shown for flow rates.}
\end{figure}

\noindent The inserts in Fig. 5 exhibit percentile variations in AF with respect to NSR, by focusing on the maximum values for pressures (Fig. 5a,c) and minimum values for flow rates (Fig. 5b,d). Although the input pressure is more likely to present hypotensive events, partially due to an averagely lower $P_a$ in AF \cite{MBEC_2014}, on the contrary the probability of hypertensive events increases along the ICA-MCA pathway, with a maximum at the capillary level (95 $\%$ in NSR corresponds to less than 70 $\%$ in AF). For flow rates we concentrate on the percentile variations of the minima, as possible quantification of hypoperfusive events. No significant differences emerge when moving from large arteries towards the deep cerebral circulation (Fig. 5, panels b, d, f, h). Contrary to hypertensive events, which are mainly linked to the instantaneous maximum pressure values reached, hypoperfusions are more related to the temporal persistence of the flow rate below the physiological thresholds \cite{SciRep_2016}. Therefore, pressure maxima are indicators of increased hypertensive events, while flow rate minima - being markers of low instantaneous flow rate - are not analogously symptomatic of hypoperfusions. An improved interpretation can be gained through the analysis of mean values per beat, which is offered in the following sections.

\subsubsection{Analysis of mean values per beat}
\begin{figure}[h!]
\includegraphics[width=\columnwidth]{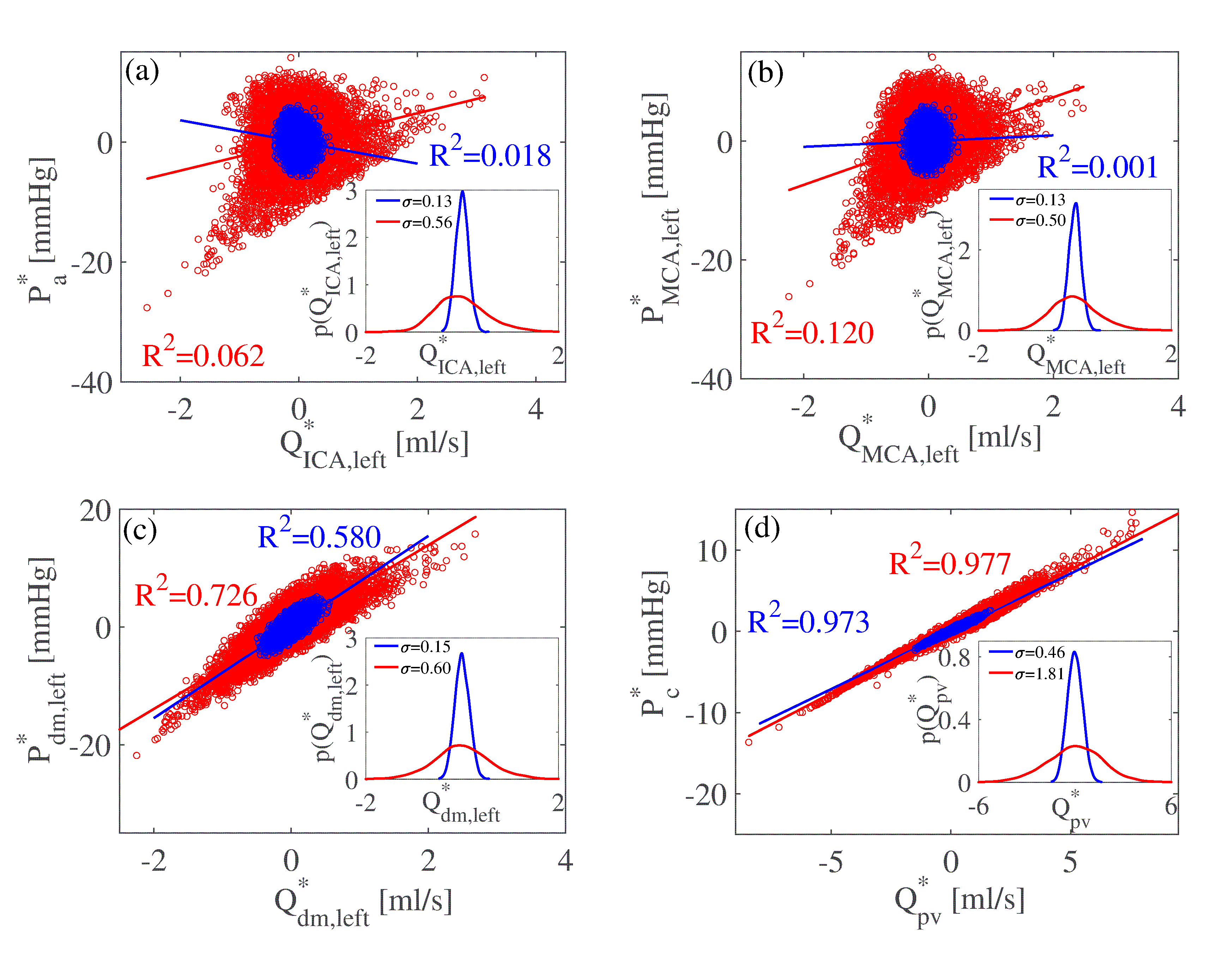}
\caption{Analysis of the mean values per beat (blue: NSR, red: AF). Scatter plots: (a) large arteries $P^*_a$ and $Q^*_{ICA,left}$; (b) middle cerebral artery $P^*_{MCA,left}$ and $Q^*_{MCA,left}$; (c) middle distal region $P^*_{dm,left}$ and $Q^*_{dm,left}$; (d) capillary-venous region $P^*_c$ and $Q^*_{pv}$. Inserts represent the PDFs of mean flow rate values, $Q^*_i$. Coefficients of determination, $R^2$, and the linear fittings are computed for each configuration.}
\end{figure}
The mean values per beat are computed for pressure ($\overline{P_i}$) and flow rate ($\overline{Q_i}$) over the 5000 cardiac periods ($i=1,...,5000$). These 5000 values are referred to the mean values, $\overline{P}$ and $\overline{Q}$, of the complete temporal signals: $P^*_i = \overline{P}_i - \overline{P}$ and $Q^*_i = \overline{Q}_i - \overline{Q}$. In Fig. 6, flow rate-pressure scatter plots are reported in NSR and AF conditions for the internal carotid artery (panel a), middle cerebral artery (panel b), middle distal region (panel c), and the capillary-venous region (panel d), together with a linear regression data fitting for each condition with the corresponding coefficient of determination, $R^2$. Data dispersion is high at the large arteries level ($R^2<0.1$), while it decreases towards the micro-circulation, reaching $R^2$ values around 0.97, with a strict direct proportionality between $Q_{pv}^*$ and $P_c^*$. This implies that, at the capillary-venous level, hypertensive events are strictly concomitant with hyperperfusions, while hypotensive episodes occur during hypoperfusions. The present behaviour is observed in both NSR and AF; however, the range of AF data is much wider, reaching extreme values.

\noindent The inserts in Fig. 6a-d represent the PDFs of the mean flow rates, $Q^*_i$, confirming the enhanced variability going to the peripheral regions, which is 3 times higher than in the large arteries. Moreover, standard deviation values, $\sigma$, are about 4 times higher in AF than in NSR at each region. This combined increase in variability leads to extremely high data dispersion in the micro-circulation during AF and underlies the mechanisms promoting the presence of hypoperfusions.

\subsubsection{Multivariate and univariate linear regression models}

\begin{figure}[h!]
\includegraphics[width=\columnwidth]{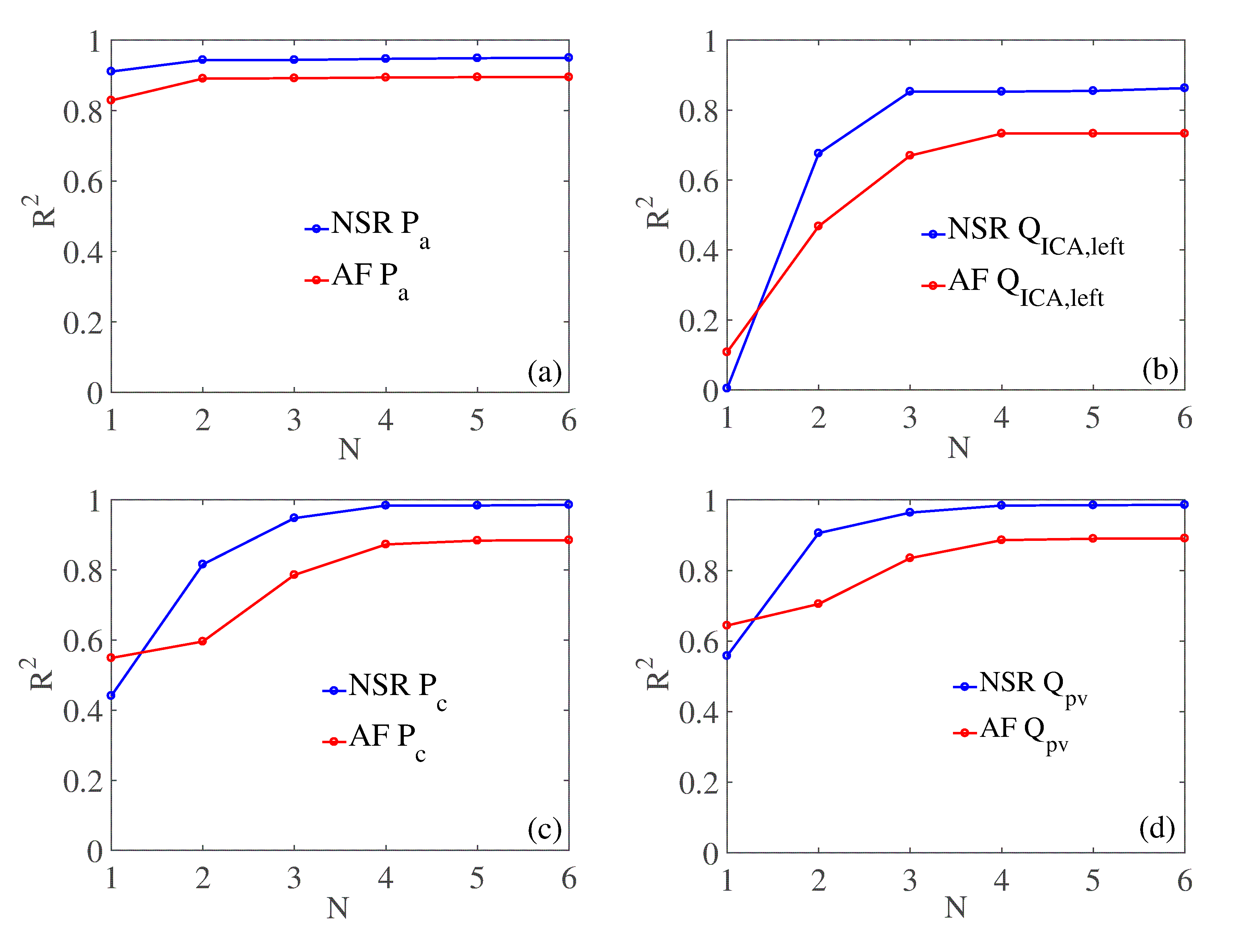}
\caption{Multivariate linear regression models in NSR (blue) and AF (red) conditions, for pressure ((a,c), $P_a$ and $P_c$) and flow rate ((b,d), $Q_{ICA,left}$ and $Q_{pv}$). Coefficients of determination, $R^2$, as a function of the number of regressors, $N$.}
\end{figure}

Multivariate linear regression models are shown here for the mean values $\overline{P}_i$ and $\overline{Q}_i$ in the four regions, having as regressors the preceding beats. The current mean value is indicated as $\overline{P}_0$ and $\overline{Q}_0$, while with $RR_{-i}$ we refer to the i-th preceding beat. The models are formalized as follows:

\begin{eqnarray}
\overline{P}_0 = a + \sum_{i=1}^N \alpha_i RR_{-i}, \\
\overline{Q}_0 = b + \sum_{i=1}^N \beta_i RR_{-i},
\end{eqnarray}

\noindent where $a$ and $b$ are the intercept values, $\alpha_i$ and $\beta_i$ represent the coefficients of the linear multivariate model, while $N$ is the number of regressors. By choosing $N$, a multivariate model is built with $N$ regressors. The number of regressors was tested up to $N=6$, leading to 6 models for pressure and 6 for flow rate, at each cerebral region and rhythm condition (96 models are computed in total). With $N=1$, the multivariate linear regression model turns into a univariate linear regression model.

In Fig. 7, the coefficients of determination, $R^2$, are shown for each model as a function of the number of regressors, $N$, in NSR and AF conditions for pressures and flow rates at the large arteries level ($P_a$ and $Q_{ICA,left}$) and in the micro-circulation ($P_c$ and $Q_{pv}$). In general, AF presents lower $R^2$ values than NSR for the multivariate models, while for the univariate models ($N=1$) AF shows higher $R^2$ values than NSR for all flow rates and capillary pressures. Moreover, in both NSR and AF, $R^2$ values are higher in the peripheral regions than the large arteries, since the signal average amplitude decreases going downstream (e.g., in AF $P_{a,max} - P_{a,min}=42$ mmHg, $P_{c,max} - P_{c,min}=6$ mmHg).

\noindent For pressures, in the large arteries region the univariate models capture most of the correlation, having $R^2>0.8$ for both NSR and AF. $RR_{-1}$ and $RR_{-2}$ are sufficient to accurately predict the current pressure level, $\overline{P}_0$, at the cerebral entrance. At the capillary level, instead, $R^2$ reaches a plateau for $N=4$, meaning that 4 preceding beats are necessary to retain the present haemodynamic content. For flow rates, in the large arteries region univariate models are not very informative ($R^2<0.2$), while the main correlation content is retained by $RR_{-2}$ and saturated with $RR_{-3}$. In the peripheral region, univariate models gain relevance exceeding $R^2=0.6$, however it is necessary to consider up to 4 regressors to guarantee a plateau for $R^2$.

\begin{figure}
\includegraphics[width=\columnwidth]{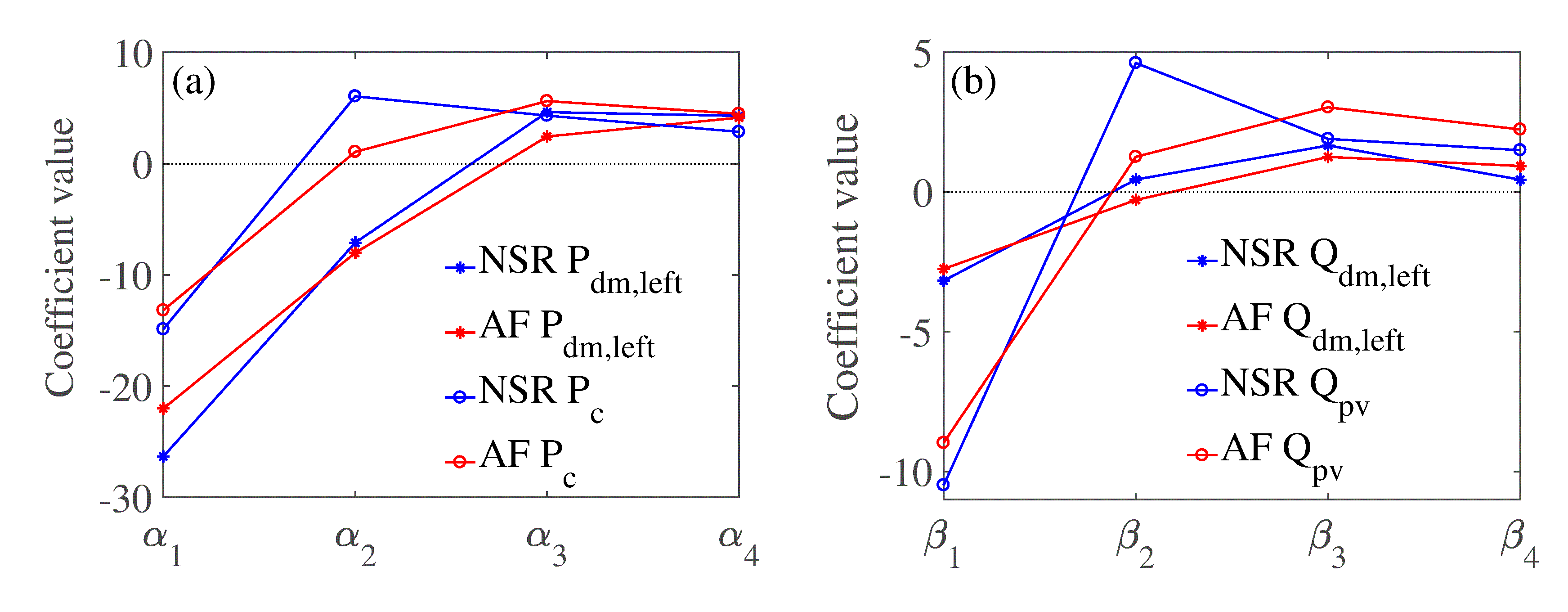}
\caption{Coefficients $\alpha_i$ and $\beta_i$ of the linear multivariate models, $i=1,...N$, with $N=4$. (a) Pressures, $P_{dm,left}$ and $P_c$; (b) flow rates, $Q_{dm,left}$ and $Q_{pv}$. NSR: blue, AF: red.}
\end{figure}

\noindent For both pressures and flow rates, the number of preceding beats necessary to fully describe the current state increases towards the micro-circulation. It should, however, be noted that, in all the regions and rhythm conditions, the multivariate model with 4 regressors ($N=4$) represents the maximum correlation level obtained. In fact, beyond this threshold, by adding further regressors the prediction of the current state is not improved. We can therefore conclude that $RR$ beats are significant regressors and the present haemodynamic state has memory of the past to the extent of about 4 beats. In other words, 4 consecutive beats are sufficient to predict the next pressure and flow rate levels.

To better explore the role and weight of the preceding beats, the multivariate models retaining the maximum correlation level (4 regressors) are now analyzed in detail, with the related coefficients $\alpha$ and $\beta$ shown in NSR and AF for the distal and capillary-venous sections (see Fig. 8). Coefficients $\alpha_1$ and $\beta_1$ are always negative with large absolute values, meaning that a substantial contribution to potential hypoperfusions and hypotensive episodes is linked to the length of $RR_{-1}$. With a long $RR_{-1}$ beating, the terms $\beta_1 RR_{-1}$ and $\alpha_1 RR_{-1}$ become predominant, leading to a low flow rate and low pressure levels. Coefficients related to $RR_{-2}$ are more variable, since $\alpha_2$ and $\beta_2$ are positive in the capillary-venous region, while in the distal region $\alpha_2$ are negative and $\beta_2$ are close to zero. Coefficients $\alpha_3$ and $\beta_3$ present moderate positive values, and this scenario is found again with no considerable variation for the $\alpha_4$ and $\beta_4$ coefficients.

\noindent Based on the regression coefficients, the most dangerous RR combinations can be finally studied, i.e. those configurations which are able to minimize flow rate (hypoperfusion) and maximize pressure (hypertensive episode). We only consider the AF condition, where the $RR$ beating is uncorrelated. In NSR, instead, the beating is correlated and closely varies around 0.8 s, therefore all coefficients have the same weight as they refer to beats which all remain strictly around 75 b.p.m.

\begin{itemize}
\item Hypoperfusions can be obtained with the following quadruplets:
\begin{itemize}
\item $Q_{dm,left}$: $RR_{-1}$ long beat, $RR_{-2}$ any beat, $RR_{-3}$ short beat, $RR_{-4}$ normal/short beat;
\item $Q_{pv}$: $RR_{-1}$ long beat, $RR_{-2}$ normal/short beat, $RR_{-3}$ short beat, $RR_{-4}$ normal/short beat.
\end{itemize}
\item Hypertensive episodes may occur with the following quadruplets:
\begin{itemize}
\item $P_{dm,left}$: $RR_{-1}$: short beat, $RR_{-2}$ short/normal beat, $RR_{-3}$ normal/long beat, $RR_{-4}$ long beat;
\item $P_c$: $RR_{-1}$ short beat, $RR_{-2}$ any beat, $RR_{-3}$ long beat, $RR_{-4}$ normal/long beat.
\end{itemize}
\end{itemize}

The least probable configuration is the distal hypertension, since a sequence of consecutive beats with decreasing duration (from long to short) has to occur, while the most probable combinations are represented by distal hypoperfusion and capillary-venous hypertension. In fact, to obtain one of these two conditions at the current state, it is sufficient to have a long (or short) last beat and a short (or long) third to last beat, which is quite a plausible circumstance in AF.

\section{Discussion}

By means of different statistical tools, the signal analysis so far described not only underlines the increasing impact of AF on the cerebral micro-circulation, but also suggests a coherent framework explaining why AF induces such evident haemodynamic changes. The key point is how differently the cerebral regions respond to the alteration of the cardiac rhythm.

\begin{figure}
\includegraphics[width=\columnwidth]{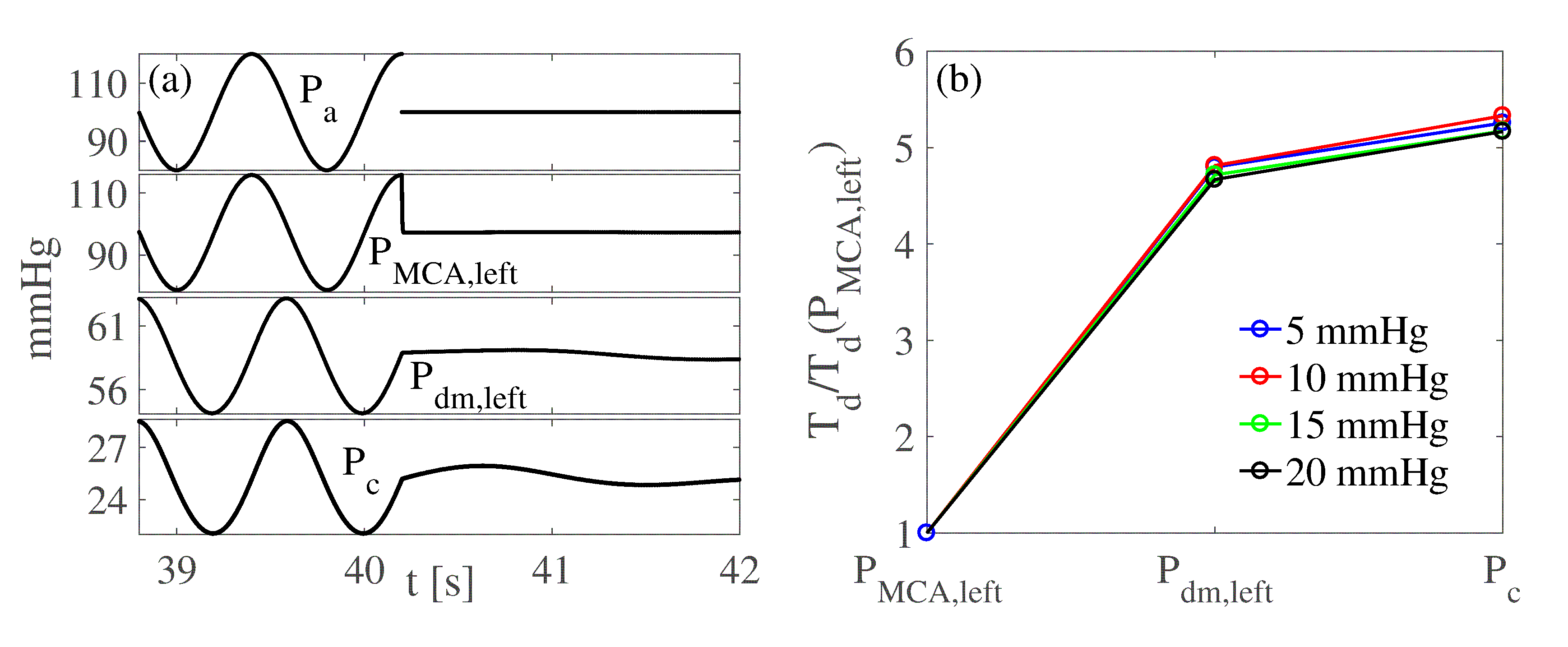}
\caption{(a) Sinusoidal $P_a$ signal (mean 100 mmHg, amplitude 20 mmHg) abruptly interrupted at its maximum (top panel) and pressure response in the downstream cerebral districts ($P_{MCA,left}$, $P_{dm,left}$, $P_c$). (b) Temporal delays, $T_d$, normalized with respect to the corresponding value in the first region ($P_{MCA,left}$) along the ICA-MCA pathway.}
\end{figure}

\noindent To better understand this, we forced the cerebral modelling with an idealized input. A sinusoidal input signal (period $T=0.8$ s) is taken for $P_a$, with mean value 100 mmHg and four different amplitudes ($100 \pm 5$ mmHg, $100 \pm 10$ mmHg, $100 \pm 15$ mmHg, $100 \pm 20$ mmHg). When at the maximum or minimum amplitude the signal is abruptly interrupted and instantaneously jumps to the mean value, then maintaining this steady state: the first case is defined as an up-mean jump (an example with an amplitude of 20 mmHg is reported in Fig. 9a, top panel), while the second represents a down-mean jump. This approach is borrowed from the theory of dynamical systems, where the system is excited by an external impulsive forcing to understand its response time. In this case, the system involved is the cerebral circulation and reacts in the different downstream districts as reported in Fig. 9a (from top, $P_a$, to bottom, $P_c$). Two basic remarks arise: (i) due to the inertia of the system, the signal in the downstream sections does not immediately reach the steady level, but it goes on oscillating with a damped amplitude before recovering the equilibrium state, and (ii) the transient damping behaviour considerably varies along the ICA-MCA pathway. To associate a quantitative measure with the transient dynamics, in every district downstream from the carotid entrance we evaluate the time lapse, $T_d$, necessary to reach the steady constant levels of pressure. $T_d$ represents the temporal delay or latency to recover the equilibrium constant state in response to a sudden and abrupt variation, and it is identified by $|\textmd{d}P/\textmd{d}t<\epsilon|$ (here $\epsilon=10^{-7}$) at each section. The latency $T_d$ at the entrance ($P_a$ level) is 0, since the jump is instantaneous, while it has a finite value immediately downstream. In Fig. 9b, we report the time delay for the pressure over the ICA-MCA pathway normalized with respect to the corresponding value at the first region, $P_{MCA,left}$. For each of the four jump amplitudes (5 mmHg, 10 mmHg, 15 mmHg, 20 mmHg), the two up-mean and down-mean jumps give similar results, thus the average value between them is taken. Since absolute values of $T_d$ depend on the threshold $\epsilon$ and the jump amplitude, we consider the time delay normalized with respect to the upstream district. In fact, the focus is not on the specific value assumed by the latency but on its variation towards the micro-circulation.

It can be noted that the normalized behaviours of $T_d/T_d(P_{MCA,left})$ do not practically depend on the jump amplitude. In the micro-circulation the latency in recovering the equilibrium state is about five times greater than that at the beginning of the middle cerebral artery. The longer delay is due to the interplay between the different mechanical features of the cerebral system, which here is modelled as an electric circuit, composed by a network of resistances and compliances (Panel 3 of Fig. 1). These mechanical and structural properties make the inertia of the system increase when entering the cerebral circulation towards the micro-vasculature. As a consequence, when a disturbance at the carotid level propagates into the cerebral vessel network, the distal and capillary regions remain altered for longer. The behaviour is analogous to that of a system of springs in series and parallel, which is externally excited at one end: each spring stiffness combines with the others, and, at a point far from the perturbed end, the damping of the oscillation is lengthened, even if the external perturbation is ceased.

The synthetic alteration of the carotid signal here described is a limit case, but it relates well to the fundamental mechanism underlying the results described in the above signal analysis. In fact, AF leads to an irregular RR series, which in turn promotes - through the systemic circulation - a collection of in-series pressure disturbances at the carotid level. Each of these perturbations singularly produces an alteration of the cerebral circulation. The higher mechanical inertia in the peripheral districts explains why here during AF right tails of the crossing time, $T_{cr}$, become important (Section 3.1.2). In fact, every $P_a$ modification leads to a downstream signal excursion from the physiological threshold (i.e., mean value in NSR). The consecutive time lapse spent above or below this threshold allows the signal to reach maximum, mean or minimum values which definitely exceed the NSR range (Sections 3.2.1 and 3.2.2). When the signal is uninterruptedly above the threshold, hyperperfusions and hypertensive events are promoted. When the contrary holds, hypotensive episodes and hypoperfusions occur.

The continuous sequence of transient perturbations at the carotid entrance represented by the AF beating does not allow the system to recover the physiological state before another disturbance arrives. The uncorrelated nature of AF beating enhances the complexity of the deep cerebral signal and reduces its predictability, since a disturbance can lead the system to the same or opposite direction with respect to the previous perturbation. As a consequence, the signal periodicity breaks towards the micro-circulation, provoking a decrease in the correlation and a reduction in the coherence time (Section 3.1.1). Although the predictive grade remains satisfactory, the increased signal complexity and uncertainty make the regression models perform less well in AF than in NSR (Section 3.2.3). Up to three or four preceding beats are necessary to averagely describe the current haemodynamic state and this temporal range is governed by the combined interplay between the superposition of different transient disturbances introduced into the system by the AF and the intrinsic latency of the system (Section 3.2.3). However, characterization of the present haemodynamic state in the first cerebral regions requires in general fewer preceding beats than in the capillary regions, that is, signal predictability deteriorates towards the peripheral regions. This aspect furthermore strengthens the basic mechanism described throughout the Discussion. Due to the mechanical features of the different cerebral districts and their reciprocal interconnection, the micro-circulation suffers much more and for longer from the AF-induced haemodynamic alterations.

\subsection{Limitations}

The limiting aspects of the present work are related to the computational hypotheses. The modelling of the cardiovascular system providing the pressure input, $P_a$, for the cerebral dynamics does not account for the baroreceptor mechanisms in the short-term. Moreover, AF is simulated assuming an uncorrelated beating and no atrial contraction, but with no increase in the constant baseline value of elastance with respect to NSR. Additionally, no long-term remodelling effects are captured and no reduced ventricular contractility is assumed. In the cerebral modelling, NSR and AF configurations solely differ by the entrance inputs, $P_a$, while the remaining haemodynamic framework is set as in healthy conditions.

\section{Conclusions}

Several haemodynamic mechanisms have been recently proposed for the association between AF and cognitive dysfunction independent of clinically relevant
events. However, definitive clinical evidence is still missing, and, at the present stage, an \emph{in silico} approach can be valuable in providing and addressing new haemodynamic-based suggestions for primary medical treatments.

\noindent Through an accurate and diversified signal analysis the present work shows a range of possible symptoms for the alteration of the haemodynamic patterns during AF in the cerebral micro-circulation. AF signals in the distal-capillary circulation lose their temporal interdependence and predictability, becoming more complex and revealing short-memory features. The crossing time analysis displays an increased probability of extreme value events which, through the beat-by-beat analysis, results in hypertensive and hypoperfusive episodes. The RR beats turn out to be good haemodynamic regressors. In particular, the role of the preceding beat in the current vascular state is crucial, while up to four consecutive RR beats are necessary to fully describe the averaged haemodynamic level of the next beat. Exploiting this outcome, the worst haemodynamic configuration occurs with a long (or short) last beat and a short (or long) third to last beat, which is rather common during AF \cite{Hennig_2006,Hayano_1997,Bootsma}. The intrinsic structural latency revealed by the cerebral circulation is plausibly disturbed by AF and exacerbates the observed scenario.

\noindent The framework here described can offer physically-based hints explaining why critical events, such as hypertensive or hypoperfusive episodes, are more likely to occur in the cerebral peripheral regions during AF, thereby further strengthening the haemodynamic link between AF and cognitive decline.

\section*{Authors' contributions}

SS and LR designed the study, developed the mathematical and computational models, performed the statistical analyses, and drafted the paper. All authors conceived the study, contributed to the interpretation of the data and manuscript writing, and gave final approval for publication.

\section*{Appendix A. Autocorrelation function, $R(\tau)$, and coherence time, $\tau_c$}

\renewcommand{\theequation}{A\arabic{equation}}
\setcounter{equation}{0}

The autocorrelation function, $R(\tau)$, representing the correlation of the signal with itself at different temporal lags $\tau$, detects repeating temporal patterns and periodicity. Through its envelope, $R_{env}(\tau)$, the autocorrelation function is used as a standard measure of the coherence time, $\tau_c$ \cite{Shin_2008}:

\begin{equation}
\tau_c = \int_{-\infty}^{+\infty} |R_{env}(\tau)|^2 d \tau
\end{equation}

The coherence time, $\tau_c$, quantifies the degree of temporal correlation of the signal: long-term coherent signals have autocorrelation functions with a slow rate of decay, while short-term memory signals (such as random signals) show very rapidly decaying autocorrelation functions. Within bioelectromagnetic signals, long-term refers to coherence times equal to or greater than 1-2 s \cite{Shin_2008}. Coherence time values, $\tau_c$, are reported in Table 2, for both NSR and AF conditions along the ICA-MCA pathway.

\begin{table}[h!]
\centering
\begin{tabular}{|c|c|c|}
  \hline
   & NSR & AF \\
   \hline
  $P_a$ & 3.09 s  & 1.11 s \\
   \hline
  $P_{c}$ & 3.29 s  & 0.90 s  \\
   \hline
  $Q_{ICA,left}$ &  3.02 s  & 0.94 s  \\
   \hline
  $Q_{pv}$ &  3.47 s  &  0.87 s \\
  \hline
\end{tabular}
\label{coherence_table}
\caption{Coherehce times, $\tau_c$, for the haemodynamic signals in the large arteries ($P_a$ and $Q_{ICA,left}$) and in the capillary/venous region ($P_c$ and $Q_{pv}$), for NSR (II column) and AF (III column).}
\end{table}

\section*{Appendix B. Minimum and maximum values analysis}

Table 3 presents the mean and standard deviation values of the 5000 minimum and maximum values for the haemodynamic variables.

\begin{table}[h!]
\hspace{-0.8cm}
\begin{tabular}{|c|c|c|c|c|}
  \hline
    & \multicolumn{2}{|c|}{NSR} & \multicolumn{2}{|c|}{AF}\\
  \hline
  Variable & Minimum & Maximum & Minimum & Maximum\\
  \hline
  $P_a$ [mmHg] & 77.91 $\pm$ 2.89 & 122.44 $\pm$ 1.55 & 75.34 $\pm$ 8.48 & 117.72 $\pm$ 4.46 \\
  \hline
  $P_{MCA,left}$ [mmHg] & 76.59 $\pm$ 2.82 & 118.25 $\pm$ 1.58 & 73.96 $\pm$ 8.22 & 113.48 $\pm$ 4.40 \\
  \hline
  $P_{dm,left}$ [mmHg] & 53.38 $\pm$ 1.63 & 61.57 $\pm$ 1.23 & 51.70 $\pm$ 5.00 & 59.60 $\pm$ 4.11 \\
  \hline
  $P_c$ [mmHg] & 21.69 $\pm$ 0.68 & 27.40 $\pm$ 0.67 & 21.93 $\pm$ 2.71 & 27.63 $\pm$ 2.76 \\
  \hline
  $Q_{ICA,left}$ [ml/s] & 2.25 $\pm$ 0.13 & 7.38 $\pm$ 0.21 & 2.35 $\pm$ 0.53 & 7.50 $\pm$ 0.78 \\
  \hline
  $Q_{MCA,left}$ [ml/s] & 1.90 $\pm$ 0.12 & 5.84 $\pm$ 0.19 & 1.96 $\pm$ 0.46 & 5.89 $\pm$ 0.68 \\
  \hline
  $Q_{dm,left}$ [ml/s] & 2.99 $\pm$ 0.15 & 4.28 $\pm$ 0.16 & 3.04 $\pm$ 0.57 & 4.31 $\pm$ 0.59 \\
  \hline
  $Q_{pv}$ [ml/s] & 9.70 $\pm$ 0.49 & 14.49 $\pm$ 0.43 & 9.96 $\pm$ 1.78 & 14.58 $\pm$ 1.69 \\
  \hline
\end{tabular}
\caption{Mean and standard deviation values, in NSR and AF conditions, of the maxima and minima of pressures ($P_{max}$ and $P_{min}$) and flow rates ($Q_{max}$ and $Q_{min}$).}
\end{table}


\begin{thebibliography}{9}

\bibitem{Piccini_2014} Piccini JP, Daubert JP 2014 Atrial fibrillation and sudden cardiac death: is heart failure the middleman? \textit{JACC: Heart Failure} \textbf{2}, 3 228–-229.

\bibitem{Lloyd-Jones_2010} Lloyd-Jones D, et al. 2010 Heart Disease and Stroke Statistics—2010 Update. \textit{Circulation} \textbf{121}, e46--e215.

\bibitem{Wolf_1991} Wolf PA, Abbott RD, Kannel WB 1991 Atrial-fibrillation as an independent risk factor for stroke - the Framingham-Study. \textit{Stroke} \textbf{22}, 8 983--988.

\bibitem{Zhu_1998} Zhu L, Fratiglioni L, Guo ZC, Aguero-Torres H, Winblad B, Viitanen M 1998 Association of stroke with dementia, cognitive impairment, and functional disability in the very old - A population-based study. \textit{Stroke} \textbf{29}, 10 2094--2099.

\bibitem{Kalantarian_2014} Kalantarian S, Ay H, Gollub RL, Lee H, Retzepi K, Mansour M, Ruskin JN 2014 Association Between Atrial Fibrillation and Silent Cerebral Infarctions A Systematic Review and Meta-analysis. \textit{Ann. Intern. Med.} \textbf{161}, 9 650--U158.

\bibitem{Gaita_2013} Gaita F, et al. 2013 Prevalence of silent cerebral ischemia in paroxysmal and persistent atrial fibrillation and correlation with cognitive function. \textit{J. Am. Coll. Cardiol.} \textbf{62}, 21 1990--1997.

\bibitem{Sabatini_2000} Sabatini T, Frisoni GB, Barbisoni P, Bellelli G, Rozzini R, Trabucchi M 2000 Atrial fibrillation and cognitive disorders in older people. \textit{J. Am. Geriatr. Soc.} \textbf{48}, 4 387--390.

\bibitem{Jacobs_2014} Jacobs V, Cutler MJ, Day JD, Bunch TJ 2014 Atrial fibrillation and dementia. \textit{Trends Cardiovas. Med.} \textbf{25}, 1 44–-51.

\bibitem{Kanmanthareddy} Kanmanthareddy A, et al. 2014 The impact of atrial fibrillation and its treatment on dementia. \textit{Curr. Cardiol. Rep.} \textbf{16}, 519.

\bibitem{Dixit_2015} Dixit S, Shukla P 2015 Does atrial fibrillation cause dementia? \textit{Trends Cardiovas. Med.} \textbf{25}, 1 52--53.

\bibitem{Alonso_2016} Alonso A, Arenas de Larriva AP 2016 Atrial Fibrillation, Cognitive Decline And Dementia \textit{Eur. Cardiol.} \textbf{11}, 1 49–-53.

\bibitem{Kalantarian_2013} Kalantarian S, Stern TA, Mansour M, Ruskin JN 2013 Cognitive impairment associated with atrial fibrillation A Meta-analysis. \textit{Ann. Intern. Med.} \textbf{158}, 5 338.

\bibitem{Santangeli_2012} Santangeli P et al. 2012 Atrial fibrillation and the risk of incident dementia: A meta-analysis. \textit{Heart Rhythm} \textbf{9}, 1 1761.

\bibitem{Kwok_2011} Kwok CS, Loke YK, Hale R, Potter JF, Myint PK 2011 Atrial fibrillation and incidence of dementia A systematic review and meta-analysis. \textit{Neurology} \textbf{76}, 10 914--922.

\bibitem{Hui_2015} Hui DS, Morley JE, Mikolajczak PC, Lee R 2015 Atrial fibrillation: A major risk factor for cognitive decline. \textit{Am. Heart J.} \textbf{169}, 448–-456.

\bibitem{Udompanich_2013} Udompanich S, Lip GYH, Apostolakis S, Lane DA 2013 Atrial fibrillation as a risk factor for cognitive impairment: a semi-systematic review. \textit{QJM-Int. J. Med.} \textbf{106}, 9 795--802.

\bibitem{Ott_1997} Ott A, Breteler MMB, deBruyne MC, vanHarskamp F, Grobbee DE, Hofman A 1997 Atrial fibrillation and dementia in a population-based study - The Rotterdam Study. \textit{Stroke} \textbf{28}, 2 316--321.

\bibitem{Kilander_1998} Kilander L, Andren B, Nyman H, Lind L, Boberg M, Lithell H 1998 Atrial fibrillation is an independent determinant of low cognitive function - A cross-sectional study in elderly men. \textit{Stroke} \textbf{29}, 9 1816--1820.

\bibitem{Chen_2016} Chen LY et al. 2016 Persistent but not paroxysmal atrial fibrillation is independently associated with lower cognitive function. \textit{J. Am. Coll. Cardiol.} \textbf{67}, 11 1379--1380.

\bibitem{Chen_2014} Chen LY, Lopez FL, Gottesman RF, Huxley RR, Agarwal SK, Loehr L, Mosley T, Alonso A 2014 Atrial Fibrillation and Cognitive Decline-The Role of Subclinical Cerebral Infarcts The Atherosclerosis Risk in Communities Study. \textit{Stroke} \textbf{45}, 9 2568.

\bibitem{Miyasaka_2007} Miyasaka Y et al. 2007 Risk of dementia in stroke-free patients diagnosed with atrial fibrillation: data from a community-based cohort. \textit{Eur. Heart J.} \textbf{28}, 16 1962--1967.

\bibitem{Dublin_2011} Dublin S et al. 2011 Atrial fibrillation and risk of dementia: A Prospective Cohort Study. \textit{J. Am. Geriatr. Soc.} \textbf{59}, 8 1369--1375.

\bibitem{Thacker_2013} Thacker EL, et al. 2013 Atrial fibrillation and cognitive decline A longitudinal cohort study. \textit{Neurology} \textbf{81}, 2 119--125.

\bibitem{Bunch_2010} Bunch et al. 2010 Atrial fibrillation is independently associated with senile, vascular, and Alzheimer's dementia. \textit{Heart Rhythm} \textbf{7}, 4 433--437.

\bibitem{Marzona_2012} Marzona I, O'Donnell M, Teo K, Gao P, Anderson C, Bosch J, Yusuf S 2012 Increased risk of cognitive and functional decline in patients with atrial fibrillation: results of the ONTARGET and TRANSCEND studies. \textit{Can. Med. Assoc. J.} \textbf{184}, 6 E329--E336.

\bibitem{O'Connell_1998} O'Connell JE, Gray CS, French JM, Robertson IH 1998 Atrial fibrillation and cognitive function: case-control study. \textit{J. Neurol. Neurosurg. Psychiatry} \textbf{65}, 3 386--389.

\bibitem{Park_2007} Park H, Hildreth A, Thomson R, O'Connell J 2007 Non-valvular atrial fibrillation and cognitive decline: a longitudinal cohort study. \textit{Age Ageing} \textbf{36}, 2 157--163.

\bibitem{Rastas_2007} Rastas S, Verkkoniemi A, Polvikoski T, Juva K, Niinisto L, Mattila K, Lansimies E, Pirttila T, Sulkava R 2007 Atrial fibrillation, stroke, and cognition - A longitudinal population-based study of people aged 85 and older. \textit{Stroke} \textbf{38}, 5 1454--1460.

\bibitem{Flaker_2010} Flaker GC, Pogue J, Yusuf S, Pfeffer MA, Goldhaber SZ, Granger CB, Anand IS, Hart R, Connolly SJ 2010 Cognitive Function and Anticoagulation Control in Patients With Atrial Fibrillation. \textit{Circ Cardiovasc Quality Outcomes} \textbf{3}, 3 277--U87.

\bibitem{Bunch_2011} Bunch TJ et al. 2011 Patients Treated with Catheter Ablation for Atrial Fibrillation Have Long-Term Rates of Death, Stroke, and Dementia Similar to Patients Without Atrial Fibrillation. \textit{J. Cardiovasc. Electrophysiol.} \textbf{22}, 8  839--845.

\bibitem{Kalantarian_2016} Kalantarian S, Ruskin JN 2016 Atrial Fibrillation and Cognitive Decline Phenomenon or Epiphenomenon? \textit{Cardiol. Clin.} \textbf{34}, 2  279.

\bibitem{Sadraie_2014} Sadraie SH, Abdi M, Navidbakhsh M, Hassani K, Kaka GR 2014 Modeling the Heart Beat, Circle of Willis and Related Cerebral Stenosis Using an Equivalent Electronic Circuit. \textit{Biomed. Eng.-App. Bas. C.} \textbf{26}, 5 1450052.

\bibitem{Campodeano_2015} Campo-Deano L, Oliveira MSN, Pinho FT 2015  A Review of Computational Haemodynamics in Middle Cerebral Aneurysms and Rheological Models for Blood Flow. \textit{Appl. Mech. Rev.} \textbf{67}, 3 030801.

\bibitem{SciRep_2016} Anselmino M, Scarsoglio S, Saglietto A, Gaita F, Ridolfi L. 2016 Transient cerebral hypoperfusion and hypertensive events during atrial fibrillation: a plausible mechanism for cognitive impairment. \textit{Sci. Rep.} \textbf{6}, 28635.

\bibitem{MBEC_2014} Scarsoglio S, Guala A, Camporeale C, Ridolfi L. 2014 Impact of atrial fibrillation on the cardiovascular system through a lumped-parameter approach. \textit{Med. Biol. Eng. Comput.} \textbf{52}, 905--920.

\bibitem{Hennig_2006} Hennig T, Maass P, Hayano J, Heinrichs S 2006 Exponential distribution of long heart beat intervals during atrial fibrillation and their relevance for white noise behaviour in power spectrum. \textit{J. Biol. Phys.} \textbf{32}, 383–-392.

\bibitem{Hayano_1997} Hayano J, Yamasaki F, Sakata S, Okada A, Mukai S, Fujinami T 1997  Spectral characteristics of ventricular response to atrial fibrillation. \textit{Am. J. Physiol.-Heart C} \textbf{273}, 6 H2811--H2816.

\bibitem{Shi_2006} Korakianitis T, Shi Y 2006 Numerical simulation of cardiovascular dynamics with healthy and diseased heart valves. \textit{J. Biomech.} \textbf{39} 1964–-1982.

\bibitem{CMBBE_2016} Scarsoglio S, Camporeale C, Guala A, Ridolfi L. 2016 Fluid dynamics of heart valves during atrial fibrillation: a lumped parameter-based approach. \textit{Comput. Methods Biomech. Biomed. Engin.} \textbf{10}, 19 1060–-1068.

\bibitem{PeerJ_2016} Scarsoglio S, Saglietto A, Gaita F, Ridolfi L, Anselmino M. 2016 Computational fluid dynamics modelling of left valvular heart diseases during atrial fibrillation. \textit{PeerJ} \textbf{4}, e2240.

\bibitem{PlosOne_2015} Anselmino M, Scarsoglio S, Saglietto A, Gaita F, Ridolfi L. 2015 Rate control management of atrial fibrillation: may a mathematical model suggest an ideal heart rate? \textit{PLoS ONE} \textbf{10}, 3 e119868.

\bibitem{PlosOne_2017} Anselmino M, Scarsoglio S, Saglietto A, Gaita F, Ridolfi L. 2017 A computational study on the relation between resting heart rate and atrial fibrillation haemodynamics under exercise. \textit{PLoS ONE} \textbf{12}, 1  e0169967.

\bibitem{Ursino} Ursino M, Giannessi M. 2010 A model of cerebrovascular reactivity including the circle of Willis and cortical anastomoses. \textit{Ann. Biomed. Eng.} \textbf{38}, 955–-974.

\bibitem{Shin_2008} Shin K, Hammond J 2008 Fundamentals of Signal Processing for Sound and Vibration Engineers. \textit{Wiley Publisher} 416.

\bibitem{Bootsma} Bootsma BK, Hoelen AJ, Strackee J, Meijler FL. 1970 Analysis of R-R Intervals in Patients with Atrial Fibrillation at Rest and During Exercise. \emph{Circulation} 41, 783-794.
\end{thebibliography}
\end{document}